\documentclass[conference]{IEEEtran}
\IEEEoverridecommandlockouts
\usepackage[caption=false,font=footnotesize]{subfig}
\usepackage{algorithm}
\newtheorem{Remark}{Remark}
\usepackage{setspace}

\usepackage{cite}
\usepackage{amsmath,amssymb,amsfonts}
\usepackage{algorithmic}
\usepackage{graphicx}
\usepackage{textcomp}
\usepackage{xcolor}
\def\BibTeX{{\rm B\kern-.05em{\sc i\kern-.025em b}\kern-.08em
    T\kern-.1667em\lower.7ex\hbox{E}\kern-.125emX}}

\begin{document}

\title{
	Linear Dynamic Range Extension for Rydberg Atomic Receiver via Band-Limited Reconstruction
}

\author{
	\IEEEauthorblockN{
		Meng Fang\textsuperscript{1},
		Lebin Chen\textsuperscript{1},
		Jun-Rong Chen\textsuperscript{2},
		Ming-Min Zhao\textsuperscript{1}, 
		Bin-Bin Wei\textsuperscript{3}
	}
	\IEEEauthorblockA{
		\textsuperscript{1}College of Information Science and Electronic Engineering, Zhejiang University, Hangzhou 310027, China\\
		Zhejiang Provincial Key Laboratory of Multi-Modal Communication Networks and Intelligent Information Processing\\
		National Key Laboratory of Millimeter-Wave and Terahertz Remote Sensing, Hangzhou 310027, China\\
		\textsuperscript{2}School of Physics, Harbin Institute of Technology, Harbin 150001, China\\
		\textsuperscript{3}Institute of System Engineering, Tianjin 300161, China\\
		Email: \{12531109, 12431101, zmmblack\}@zju.edu.cn,
		24B911010@stu.hit.edu.cn, weibb.2009@tsinghua.org.cn
	}
}

\maketitle

\begin{abstract}
	Rydberg-atom receiver offers a promising approach to radio-frequency (RF) signal reception in wireless communication and sensing systems, combining high sensitivity with the potential for broad frequency coverage.  
	However, the limited linear dynamic range (LDR) of the atomic system causes nonlinear distortion at high input signal levels, thereby restricting the receiver's effective linear input range. 
	To address this issue, we analyze the mechanisms underlying nonlinear distortion in Rydberg atomic receivers and propose a physics-informed band-limited iterative reconstruction (PI-BLIR) method. 
	Specifically, by exploiting the known nonlinear atomic response and signal bandwidth prior, we formulate a constrained maximum-likelihood problem to reconstruct the distorted complex intermediate-frequency (IF) waveform. 
	With the bandwidth constraint absorbed into the frequency-domain parameterization, the constrained maximum-likelihood problem is transformed into an unconstrained optimization problem, which is solved using gradient descent.
	The proposed method relies solely on digital post-processing and requires no changes to the local-oscillator (LO) field strength, optical parameters, or atomic operating conditions. 
	Simulation results demonstrate that PI-BLIR extends the receiver's LDR by more than 14 dB and enhances the demodulation performance of the atomic receiver.
\end{abstract}


\vspace{-4pt}
\section{Introduction}
\vspace{-4pt}

Rydberg-atom-based receivers have attracted growing attention as a promising platform for wireless communication and sensing~\cite{ChenYuanbin2026,YuanJinpeng2023}. 
Owing to their large electric dipole moments and high polarizabilities, Rydberg atoms interact strongly with radio-frequency (RF) fields, enabling microwave signals to be coupled to atomic states and read out optically~\cite{CuiMingyao2025}.
In particular, sensors based on electromagnetically induced transparency (EIT) and Autler--Townes (AT) splitting enable electric-field measurements from DC to THz \cite{Fleischhauer2005,Sedlacek2012}. 
In a Rydberg-atom superheterodyne receiver, a local-oscillator (LO) field combines with the RF signal in the atomic system to enable intermediate-frequency (IF) readout, providing high sensitivity at the $\mathrm{nV \cdot cm^{-1} \cdot Hz^{-1/2}}$ level and access to signal amplitude, frequency, and phase \cite{Jing2020,Fancher2021,Berweger2023}.

For wireless communications and radar sensing, 
linear dynamic range (LDR), alongside sensitivity and bandwidth, is a key performance metric \cite{WuFengchuan2023}. Beyond the atom's finite linear-response range, increasing the input amplitude induces nonlinear distortion and waveform degradation \cite{Goncalves2025,XiaYifan2024}, limiting wide-dynamic-range reception. Therefore, understanding nonlinear distortion mechanisms and extending LDR are essential.

Existing studies on nonlinear distortion in Rydberg atomic receivers have focused on distortion characterization and system parameter optimization. For example, the work~\cite{Goncalves2025} experimentally characterized harmonic and intermodulation distortion using single- and two-tone tests, with metrics including compression points, intermodulation intercept points, and spurious-free dynamic range. 
The work~\cite{WuFengchuan2023} modeled the response of superheterodyne receivers, analyzed harmonic generation, and investigated the effect of the probe-field Rabi frequency on the LDR. 
For amplitude modulation (AM) demodulation, other studies have reduced total harmonic distortion (THD) or improved LDR by optimizing the LO operating point, coupling-field parameters, and microwave-field parameters ~\cite{JinYuankai2023,RaoRuijian2022,WuYuhao2025}. 
Furthermore, the work\cite{QuChuan2026} used particle swarm optimization to jointly optimize the system operating parameters for the linear gain and LDR of a balanced homodyne receiver, with THD as the evaluation metric.

Beyond physical parameter optimization, digital post-processing has also been used to mitigate nonlinear distortion in Rydberg atomic receivers. 
Specifically, the work~\cite{XiaYifan2024} proposed a function-fitting-based digital compensation method for quadrature amplitude modulation (QAM) signals. 
The authors in~\cite{ChenMinze2025} modeled the nonlinear response of a Rydberg atomic homodyne receiver and developed a tailored inverse-function-based compensation method, extending the LDR by over 7 dB. 
These studies demonstrate that receiver linearity can be improved through either front-end operating-parameters optimization or digital compensation.

However, despite these advances, existing methods still suffer from several limitations. 
First, physical parameter optimization requires tuning the probe- or coupling-field Rabi frequency or the LO operating point, potentially compromising bandwidth or detection sensitivity. 
Second, inverse-function-based compensation is susceptible to noise amplification and numerical ill-conditioning. 
Therefore, it is necessary to develop a digital post-processing method that operates without altering the system's physical parameters and circumvents these two limitations.
To this end, this paper analyzes the nonlinear distortion mechanisms in Rydberg atomic receivers and proposes a physics-informed band-limited iterative reconstruction (PI-BLIR) method. 
By exploiting the known atomic response and signal bandwidth prior, the proposed method reconstructs distorted waveforms through digital post-processing without modifying the receiver's physical parameters.
Simulation results demonstrate a 14.44 dB improvement in LDR over conventional approach, along with enhanced BER performance.

\vspace{-4pt}
\section{System Model and Problem Formulation}
\vspace{-4pt}

This section describes the receiver's operating principle, analyzes the mechanisms of nonlinear distortion, and formulates an optimization model for signal recovery.

\vspace{-4pt}
\subsection{Principle of a Rydberg-atom superheterodyne receiver}
\vspace{-4pt}

\begin{figure}[!t]
	\centering
	\includegraphics[width=0.65\columnwidth]{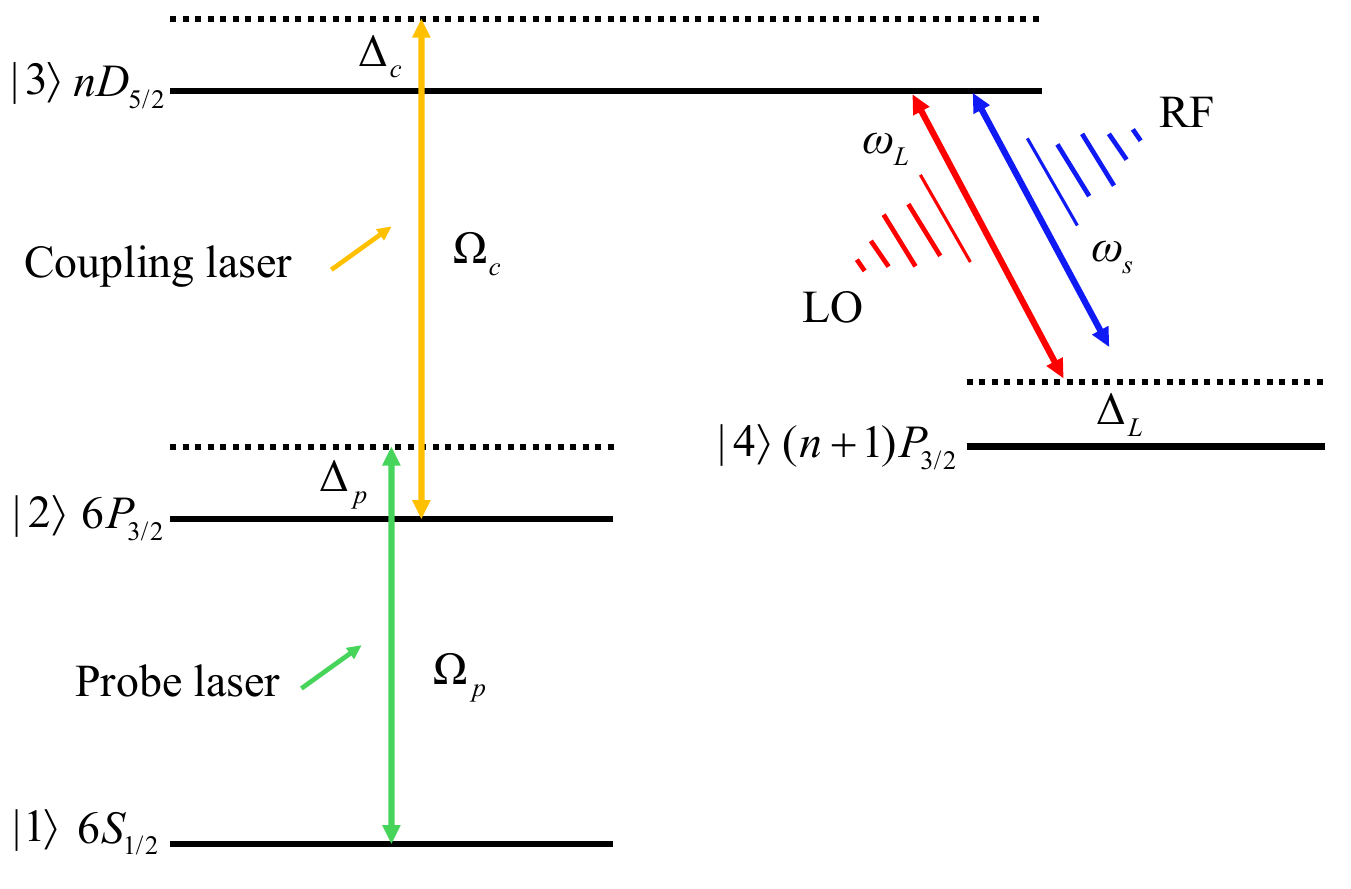}
	\caption{
		A four-level ladder-type atomic energy-level structure.
	}
	\label{fig:energy_level}
\end{figure}
\begin{figure}[!t]
	\centering
	\includegraphics[width=0.85\columnwidth]{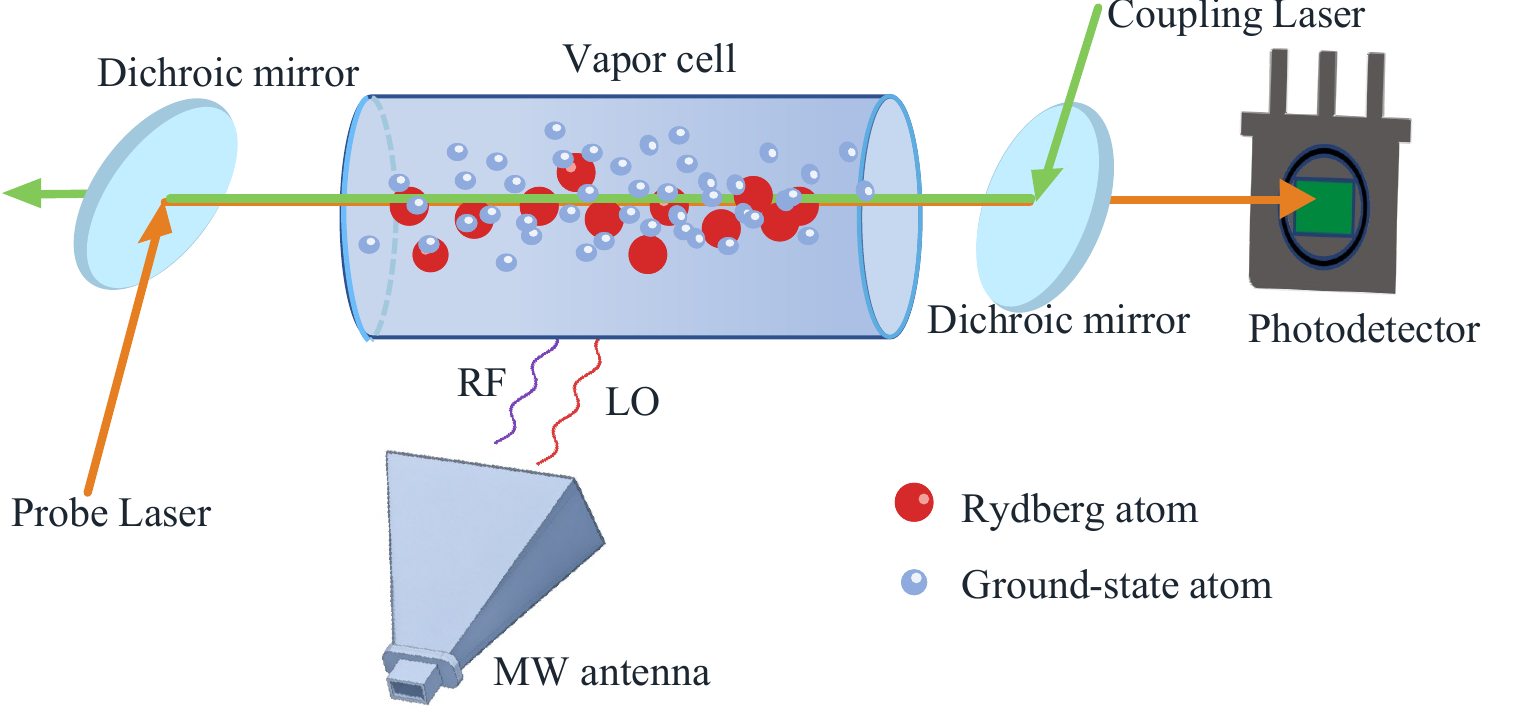}
	\caption{
		Schematic of a Rydberg-atom superheterodyne receiver.
	}
	\label{fig:receiver_schematic}
\end{figure}

In this work, we consider a two-color, four-level (2C4L) Rydberg-atom superheterodyne receiver that uses probe and coupling lasers to establish an electromagnetically induced transparency (EIT) response in an alkali vapor cell. As shown in Fig.~\ref{fig:energy_level}, the probe and coupling lasers drive the $|1\rangle\leftrightarrow|2\rangle$ and $|2\rangle\leftrightarrow|3\rangle$ transitions, with Rabi frequencies $\Omega_p$ and $\Omega_c$, respectively. The RF signal $E_{RF}$ and LO signal $E_{LO}$ jointly drive the $|3\rangle\leftrightarrow|4\rangle$ transition. As illustrated in Fig.~\ref{fig:receiver_schematic}, their complex electric-field superposition modulates the atomic response and transmitted probe power. A photodetector (PD) converts the transmitted light into an electrical signal, from which the intermediate-frequency (IF) component is extracted to detect the RF field.

The Hamiltonian of the 2C4L system is given by\cite{ChenMinze2025}
\begingroup
\setlength{\abovedisplayskip}{2pt}
\setlength{\belowdisplayskip}{2pt}
\setlength{\abovedisplayshortskip}{2pt}
\setlength{\belowdisplayshortskip}{2pt}
\[
\resizebox{\columnwidth}{!}{$
	H=\frac{\hbar}{2}
	\begin{pmatrix}
		0 & \Omega_p & 0 & 0 \\
		\Omega_p^* & -2\Delta_p & \Omega_c & 0 \\
		0 & \Omega_c^* & -2(\Delta_p+\Delta_c) & \Omega_{\mathrm{tot}} \\
		0 & 0 & \Omega_{\mathrm{tot}}^* &
		-2(\Delta_p+\Delta_c-\Delta_L)
	\end{pmatrix}.
	$}
\]
\endgroup
The detunings $\Delta_p$, $\Delta_c$, and $\Delta_L$ correspond to the probe laser, coupling laser, and LO field, respectively.
For a room-temperature atomic vapor, thermal motion
causes different velocity classes of atoms to experience
different optical detunings. 
With counter-propagating probe and coupling beams, the Doppler-shifted detunings
are $\Delta_2'=\Delta_p-2\pi v/\lambda_p$,
$\Delta_3'=\Delta_p+\Delta_c+
2\pi v(1/\lambda_c-1/\lambda_p)$, and
$\Delta_4'=\Delta_3'-\Delta_L$.
The total electric-field amplitude $E_{\mathrm{tot}}$ and the corresponding Rabi frequency $\Omega_{\mathrm{tot}}$ are related by $\Omega_{\mathrm{tot}}= \mu_{34}|E_{\mathrm{tot}}| / \hbar$, where $\mu_{34}$ is the microwave-transition dipole
moment.
Under the weak-probe approximation
($\rho_{11}\approx 1$), the steady-state probe coherence
for atoms with velocity $v$ is
\begingroup
\setlength{\abovedisplayskip}{2pt}
\setlength{\belowdisplayskip}{2pt}
\setlength{\abovedisplayshortskip}{2pt}
\setlength{\belowdisplayshortskip}{2pt}
\begin{equation}
	\rho_{21}
	=
	\frac{i\Omega_p/2}{
		(i\Delta_2'-\Gamma_{21})
		+\dfrac{|\Omega_c|^2/4}{
			(i\Delta_3'-\Gamma_{31})
			+\dfrac{|\Omega_{\mathrm{tot}}|^2/4}
			{i\Delta_4'-\Gamma_{41}}}}.
	\label{eq:rho21_doppler}
\end{equation}
\endgroup
Here, $\Gamma_{n1}$ denotes the coherence decay rate
between levels $|n\rangle$ and $|1\rangle$. Neglecting
collisional dephasing and ground-state decay gives
$\Gamma_{n1}=\Gamma_n/2$, where $\Gamma_n$ denotes the population decay
rate of level $|n\rangle$.

The atomic vapor contains atoms with a distribution
of thermal velocities. 
Therefore, the macroscopic atomic coherence, denoted by the Doppler-averaged coherence $\rho_{21}'$, is obtained by averaging the individual atomic coherences over the velocity distribution.
The optical susceptibility is then given by
$\chi=-\frac{2N_a|\mu_{12}|^2}{\varepsilon_0\hbar\Omega_p}\rho_{21}'$, where $N_a$ is the atomic number density, $\mu_{12}$ is the probe-transition dipole moment, and $\varepsilon_0$ is the vacuum permittivity.
According to the Beer-Lambert law, the probe power attenuates exponentially during propagation through the atomic vapor, with the absorption governed by the imaginary part of the susceptibility. The transmitted probe power is given by~\cite{JinYuankai2023}
\begingroup
\setlength{\abovedisplayskip}{2pt}
\setlength{\belowdisplayskip}{2pt}
\setlength{\abovedisplayshortskip}{2pt}
\setlength{\belowdisplayshortskip}{2pt}
\begin{equation}
	P_{T}=
	P_0\exp\left[
	-\frac{2\pi L}{\lambda_p}
	\operatorname{Im}\bigl(\chi\bigr)
	\right].
	\label{eq:thermal_atomic_response}
\end{equation}
\endgroup
Here, $P_0$ is incident probe power, $L$ is vapor-cell length, and $\lambda_p$ is probe wavelength. 
The steady-state response from the microwave field to the PD output is defined as
\begingroup
\setlength{\abovedisplayskip}{2pt}
\setlength{\belowdisplayskip}{2pt}
\setlength{\abovedisplayshortskip}{2pt}
\setlength{\belowdisplayshortskip}{2pt}
\begin{equation}
	S(E_{\mathrm{tot}})=G_aP_T(E_{\mathrm{tot}}),
	\label{eq:PtoS}
\end{equation}
\endgroup
where $G_a$ is the effective gain converting transmitted probe power into PD output current.


\vspace{-4pt}
\subsection{Mechanism of Nonlinear Distortion}
\vspace{-4pt}

As described above, the RF field $E_{RF}$ is superposed with the LO field $E_{LO}$ to drive the atoms, while the transmitted probe power is measured by a PD. Ideally, the detector output $S(E)$ would vary linearly with the $E_{RF}$, enabling distortion-free signal reception. Nevertheless, $S(E)$ depends nonlinearly on the total electric-field amplitude $E_{tot}$, resulting in waveform distortion. This subsection investigates the generation mechanism of such nonlinear distortion.
 
Let the complex representations of the LO $E_{LO}(t)$ and RF $E_{RF}(t)$ electric fields be $A e^{j\omega_{\mathrm{LO}}t}$ and $U(t)=|U(t)|e^{j(\omega_{\mathrm{RF}}t+\Phi(t))}$, respectively, where $U(t)$ represents a typical band-limited communication signal (e.g., QAM) carrying both amplitude and phase information.
In the considered superheterodyne receiver, the LO and signal fields are superimposed in the atomic vapor cell, yielding $E_{tot} = A e^{j(\omega_{\mathrm{LO}}t)} + |U(t)|e^{j(\omega_{\mathrm{RF}}t+\Phi(t))}$. 
To describe the atom–field interaction, we adopt a rotating frame at the LO frequency and apply the rotating-wave approximation (RWA)\cite{Fleischhauer2005}, which neglects the rapidly oscillating counter-rotating terms. 
Taking the LO phase as the rotating reference frame, the magnitude of total electric field under the RWA is written as 
$E_{tot}=[A+|U(t)|e^{j(\omega_{\mathrm{IF}}t+\Phi(t))}]e^{j(\omega_{\mathrm{LO}}t)}$. 
In the LO rotating frame, the high-frequency carrier $\omega_{LO}$ is removed, leaving the slowly varying envelope of the superimposed field as
\begingroup
\setlength{\abovedisplayskip}{2pt}
\setlength{\belowdisplayskip}{2pt}
\setlength{\abovedisplayshortskip}{2pt}
\setlength{\belowdisplayshortskip}{2pt}
\begin{equation}
	|E_{tot}|= |A+|U(t)|e^{j(\omega_{\mathrm{IF}}t+\Phi(t))}|,
\end{equation}
\endgroup 
where $\omega_{\mathrm{IF}}=\omega_{\mathrm{RF}} - \omega_{\mathrm{LO}}=2\pi f_{\mathrm{IF}}$. Thus, the atomic vapor cell acts as a mixer, down-converting the signal field $U(t)$ to the IF via slowly varying envelope extraction. The IF signal field $|U(t)|e^{j(\omega_{\mathrm{IF}}t+\Phi(t))}$ is the actual signal of interest for the receiver backend. 
Let $u(t)$ denote the corresponding IF representation of $U(t)$ after mixing with the LO. The down-conversion shifts the carrier frequency from $\omega_{\mathrm{RF}}$ to $\omega_{\mathrm{IF}}$, while preserving the amplitude envelope and phase modulation. Thus, the down-converted signal is expressed as $u(t)=|U(t)|e^{j(\omega_{\mathrm{IF}}t+\Phi(t))}$, where $|u(t)|=|U(t)|$.

Let $r(t)$ denote the slowly varying envelope of the total electric-field magnitude, given by $r(t) = |A+u(t)|$. Then, the detector observation model can be expressed
\begingroup
\setlength{\abovedisplayskip}{2pt}
\setlength{\belowdisplayskip}{2pt}
\setlength{\abovedisplayshortskip}{2pt}
\setlength{\belowdisplayshortskip}{2pt}
\begin{equation}
	y(t)=S\!\left(|A+u(t)|\right)+n(t),
	\label{eq:complete_forward_continuous}
\end{equation}
\endgroup 
where $n(t)$ is the PD output noise. 
The model exhibits two sources of nonlinearity arising from the magnitude operation $|\cdot|$ on the combined complex field and the intrinsic nonlinearity of the atomic response $S(\cdot)$.

We first examine the nonlinearity introduced by the magnitude operation $|\cdot|$.
Expanding the square of $r(t)$,
we obtain
$
r(t) = A\sqrt{1+\frac{|u(t)|^2}{A^2} + \frac{2|u(t)|}{A}\cos(2\pi f_{\mathrm{IF}}t+\Phi(t))}.
$
Then, a second-order Taylor expansion is further applied to obtain
\begingroup
\setlength{\abovedisplayskip}{2pt}
\setlength{\belowdisplayskip}{2pt}
\setlength{\abovedisplayshortskip}{2pt}
\setlength{\belowdisplayshortskip}{2pt}
\begin{equation}
	r(t) \approx A + {|u(t)|^2}\sin^2(\theta)/{2A} + |u(t)|\cos(\theta),
	\label{eq:Taylor_expansion_r(t)}
\end{equation}
\endgroup 
where $\theta$ denotes $2\pi f_{\mathrm{IF}}t+\Phi(t)$.
In conventional approaches, when the system operates in the regime $A \gg |u(t)|$, the spurious term ${|u(t)|^2}\sin^2(\theta)/{2A}$ is negligible, yielding $r(t) \approx A + |u(t)|\cos(2\pi f_{\mathrm{IF}}t+\Phi(t))$. In this case, the envelope is simply the superposition of the LO amplitude and the IF signal field, where $A$ provides a DC bias and the IF signal term provides an AC component. Consequently, the amplitude and phase information of the RF field are down-converted and modulated onto the IF waveform.
When this condition is not satisfied, the spurious term becomes non-negligible, resulting in nonlinear distortion in the extracted IF envelope.

We next examine the nonlinearity introduced by the atomic response function $S(\cdot)$.
Let $x(t) = \operatorname{Re}\{u(t)\}$ and $z(t) = \operatorname{Im}\{u(t)\}$. 
From Eq.~\eqref{eq:Taylor_expansion_r(t)}, expanding $S(r)$ around the LO operating point $A$ gives 
$S(r) \approx S(A)+S'(A)x + \frac{S'(A)}{2A}z^2
+\frac{S''(A)}{2}x^2+\cdots$.
The presence of these higher-order terms indicates that the atomic response contains nonlinear contributions beyond the first-order term.
Thus, receiver nonlinearity arises from both envelope-magnitude formation and the curvature of the atomic response. 

\begin{figure}[!h]
	\centering
	\includegraphics[width=0.82\columnwidth]{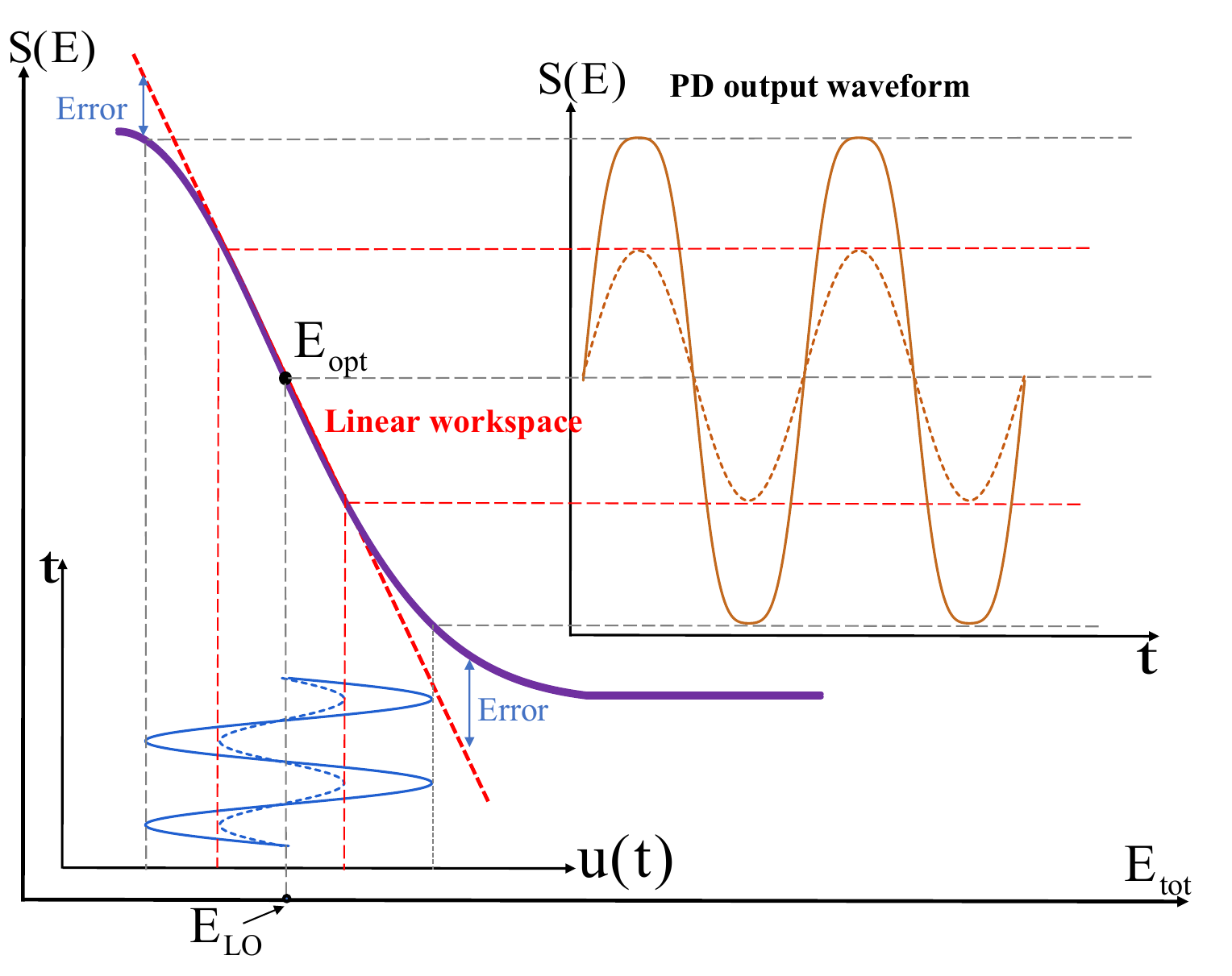}
	\caption{
		Schematic illustration of the nonlinear atomic 
		response, optimal operating point, and linear operating 
		region, based on\cite{QuChuan2026}. 
	}
	\label{fig:abab1}
\end{figure}

A schematic illustration of the nonlinear atomic response
and its approximately linear operating region is shown
in Fig.~\ref{fig:abab1}\cite{QuChuan2026}.
The blue and orange waveforms show the input signal $u(t)$
and the corresponding PD output $S(E)$, respectively.
The atomic response is approximately linear around the optimal operating point $E_{opt}$, with the area bounded by the red dashed lines denoting the linear workspace of the atomic response. Within this region, the PD output $S(E)$ varies approximately linearly with the $E_{\mathrm{tot}}$. Outside it, the envelope-magnitude operation can cause waveform degradation. 

\vspace{-4pt}
\subsection{Problem Formulation}
\vspace{-4pt}

To address the aforementioned two sources of nonlinearity, this subsection formulates the recovery of nonlinear signal distortion mathematically.
Specifically, using the observation model in~\eqref{eq:complete_forward_continuous} and the bandwidth prior of the communication signal, we formulate nonlinear distortion mitigation as a maximum-likelihood (ML) estimation problem.

Specifically, let the PD output be sampled at a sampling
rate $F_s$, with $N$ samples per frame. Define the observation vector
$\mathbf y=[y_0,\ldots,y_{N-1}]^{\mathrm T}\in\mathbb R^N$ and the
complex envelope to be reconstructed
$\mathbf u=[u_0,\ldots,u_{N-1}]^{\mathrm T}\in\mathbb C^N$,
where $y_n=y(n/F_s)$ and $u_n=u(n/F_s)$.
The pointwise nonlinear forward operator $\mathcal F(\cdot)$ is defined as
$
[\mathcal F(\mathbf u)]_n=S(|A+u_n|), n=0,\ldots,N-1.
$
The discrete observation model is therefore
$
\mathbf y=\mathcal F(\mathbf u)+\mathbf n
$,
where $\mathbf n$ denotes measurement noise.
Assuming independent Gaussian measurement noise with $\mathbf{n} \sim \mathcal{N}(0, \sigma_n^2)$, the conditional likelihood of $\mathbf y$ given $\mathbf u$ is
\begingroup
\setlength{\abovedisplayskip}{2pt}
\setlength{\belowdisplayskip}{2pt}
\setlength{\abovedisplayshortskip}{2pt}
\setlength{\belowdisplayshortskip}{2pt}
\begin{equation}
	p(\mathbf y|\mathbf u)
	=
	\frac{1}{(2\pi\sigma_n^2)^{N/2}}
	\exp\left(
	-\frac{\|\mathcal F(\mathbf u)-\mathbf y\|_2^2}
	{2\sigma_n^2}
	\right).
\end{equation}
\endgroup
Maximizing this likelihood is equivalent to minimizing
the negative log-likelihood. 
Let $\mathbb B=[f_{\mathrm L},f_{\mathrm H}]$ denote the admissible
positive-frequency IF band, where $f_{\mathrm L}$ and $f_{\mathrm H}$ denote the lower and upper frequency limits, respectively. 
Let $U(f)$ denote the spectrum of $\mathbf{u}$, and let $\operatorname{supp} U$ denote its frequency support, i.e., the set of frequencies at which $U(f)$ is nonzero.
The bandwidth prior is then expressed as $\operatorname{supp} U \subseteq \mathbb{B}$.
Incorporating the band-limited prior as a feasible-set constraint yields
the following constrained ML problem:
\begingroup
\setlength{\abovedisplayskip}{2pt}
\setlength{\belowdisplayskip}{2pt}
\setlength{\abovedisplayshortskip}{2pt}
\setlength{\belowdisplayshortskip}{2pt}
\begin{equation}
	\begin{aligned}
		\min_{\mathbf u}\quad&
		\Phi(\mathbf u)
		=\frac{1}{2}
		\|\mathcal F(\mathbf u)-\mathbf y\|_2^2\\
		\mathrm{s.t.}\quad&
		\operatorname{supp}U\subseteq\mathbb B.
	\end{aligned}
	\label{eq:piblir_optimization}
\end{equation}
\endgroup
The objective is to estimate the complex envelope $\mathbf u$ within
the prescribed bandwidth so that its predicted output best matches
the measured observations. 
Since $\mathbf u$ is a complex envelope, its discrete fourier transform (DFT) coefficients need not satisfy conjugate symmetry, and $\mathbb B$ therefore covers only the positive-frequency IF band.

\begin{Remark}
	Since the DFT-based basis transformation relies on discrete frequency bins, spectral leakage may occur.
	Common mitigation approaches include windowing, overlapping-frame processing, and guard-band extension. In this paper, we adopt guard-band extension by expanding the prescribed IF band beyond its nominal boundaries to alleviate spectral leakage. A detailed comparison of these approaches is omitted due to space limitations.
\end{Remark}

Under the noiseless observation model, unique reconstruction of the complex IF envelope is ensured if the atomic response function $S(E)$ is strictly monotonic over the total-field range and the signal-band and sampling conditions satisfy 
\begingroup
\setlength{\abovedisplayskip}{2pt}
\setlength{\belowdisplayskip}{2pt}
\setlength{\abovedisplayshortskip}{2pt}
\setlength{\belowdisplayshortskip}{2pt}
\begin{equation}
	0<f_L,\qquad f_L>f_H-f_L,\qquad F_s>2f_H.
\end{equation}
\endgroup
The proof is omitted for brevity.

\section{Band-Limited Signal Reconstruction}

This section develops an unconstrained frequency-domain formulation for band-limited signal reconstruction and solves it using gradient descent (GD) with Armijo line search.

\subsection{Unconstrained Frequency-Domain Formulation}
 
In this formulation, the bandwidth constraint is eliminated by representing the IF waveform with reduced frequency-domain coefficients.
Let $\mathbf F$ denote the normalized DFT matrix, and define 
$
	\mathbf M_{\mathbb B}
	\triangleq
	\begin{bmatrix}
		\mathbf e_{m_1},\mathbf e_{m_2},\cdots,\mathbf e_{m_K}
	\end{bmatrix}
	\in\mathbb R^{N\times K},
$
as a band selection matrix, 
where $\{m_1,\ldots,m_K\}$ are the DFT-bin indices
within the positive IF band, and $\mathbf e_{m_i}$
is the corresponding $N$-dimensional standard basis
vector.
By further defining $\mathbf{F}_{\mathbb B}\triangleq\mathbf F^{\mathrm H}\mathbf M_{\mathbb B}$, 
the reconstructed signal is expressed as $\mathbf{u}=\mathbf{F}^{\mathrm{H}}\mathbf{M}_{\mathbb B}\mathbf{c} = \mathbf{F}_{\mathbb{B}}\mathbf{c}$, where $\mathbf{c} \in \mathbb{C}^{K}$, $K$ is the number of discrete frequency points within the IF band, and $\mathbf{F}^{\mathrm{H}}$ is the inverse DFT matrix. Through this variable substitution, the bandwidth constraint is inherently absorbed into the signal model. Consequently,~\eqref{eq:piblir_optimization} can be reformulated as the following unconstrained form with respect to $\mathbf{c}$:
\begingroup
\setlength{\abovedisplayskip}{2pt}
\setlength{\belowdisplayskip}{2pt}
\setlength{\abovedisplayshortskip}{2pt}
\setlength{\belowdisplayshortskip}{2pt}
\begin{equation}
	\min_{\mathbf{c}} \widetilde{\Phi}(\mathbf{c}) = \frac12\left\| \mathcal F(\mathbf{F}_{\mathbb B}\mathbf c) - \mathbf{y} \right\|_2^2.
	\label{eq:unconstrained_frequency_objective}
\end{equation}
\endgroup

\subsection{Nonlinear ML Reconstruction}

Starting from the unconstrained optimization problem in
\eqref{eq:unconstrained_frequency_objective}, we derive the
gradient with respect to the frequency-domain coefficient
vector $\mathbf c$. At the $k$-th iteration, let
$\mathbf u_k=\mathbf F_{\mathbb B}\mathbf c_k$, and define
$
\mathbf r_k = |A+\mathbf u_k|,
\mathbf e_k = S(\mathbf r_k)-\mathbf y,
$
where the absolute value and the nonlinear response
$S(\cdot)$ are applied element-wise. In addition, define
$\mathbf q_k=\frac{A+\mathbf u_k}{|A+\mathbf u_k|}$,
and $\mathbf s_k=S'(\mathbf r_k)$,
where $\mathbf q_k$ represents the element-wise radial
direction of the complex envelope, and $\mathbf s_k$
contains the derivative of the atomic response evaluated at
$\mathbf r_k$.

For each sampling index $n=0,1,\dots,N-1$, let
$r_{k,n}=|A+u_{k,n}|$. 
For a small perturbation $\delta_{u_{k,n}}$, the first-order variation of $r_{k,n}$ is 
$\delta_{r_{k,n}} = \operatorname{Re}[q_{k,n}^{*}\,\delta_{u_{k,n}}]$.
Applying the chain rule to
$\widetilde{\Phi}(\mathbf c)$
gives
$
\delta_{\widetilde{\Phi}}
\triangleq
\operatorname{Re}[\left(\mathbf s_k\odot\mathbf e_k\odot\mathbf q_k\right)^{\mathrm H}
\delta_{\mathbf u_k}]
$,
where $\odot$ denotes element-wise multiplication. Since
$\mathbf u_k=\mathbf F_{\mathbb B}\mathbf c_k$, we have
$\delta_{\mathbf u_k}=\mathbf F_{\mathbb B}\,\delta_{\mathbf c_k}
=\mathbf F^{\mathrm H}\mathbf M_{\mathbb B}\delta_{\mathbf c_k}$.
Substituting this relation into the above expression yields
$
\delta_{\widetilde{\Phi}}
=
\operatorname{Re}\!\left\{
\left[
\mathbf M_{\mathbb B}^{\mathrm H}\mathbf F
\left(\mathbf s_k\odot\mathbf e_k\odot\mathbf q_k\right)
\right]^{\mathrm H}
\delta_{\mathbf c_k}
\right\}.
$
Therefore, the gradient of the objective function with
respect to $\mathbf c_k$ is
\begingroup
\setlength{\abovedisplayskip}{2pt}
\setlength{\belowdisplayskip}{2pt}
\setlength{\abovedisplayshortskip}{2pt}
\setlength{\belowdisplayshortskip}{2pt}
\begin{equation}
	\nabla_{\mathbf c}\widetilde{\Phi}(\mathbf c_k)
	=
	\mathbf M_{\mathbb B}^{\mathrm H}\mathbf F
	\left(\mathbf s_k\odot\mathbf e_k\odot\mathbf q_k\right).
	\label{eq:gd_gradient}
\end{equation}
\endgroup

At the $k$-th iteration, the search direction is set to
$\mathbf d_k=-\mathbf g_k$, where
$\mathbf g_k=\nabla_{\mathbf c}\widetilde{\Phi}(\mathbf c_k)$.
Starting from an initial step size $\alpha_0>0$, the
candidate step size is reduced by a factor
$\rho\in(0,1)$ until the Armijo sufficient decrease
condition is satisfied
$
\widetilde{\Phi}(\mathbf c_k-\alpha_k\mathbf g_k)
\leq
\widetilde{\Phi}(\mathbf c_k)
-\mu\alpha_k\|\mathbf g_k\|_2^2,
$
where $\mu\in(0,1)$ is the Armijo parameter.
If this condition is not met, the step size is progressively reduced according to
$
\alpha_k\leftarrow\rho\alpha_k,~0<\rho<1, 
$
until the sufficient decrease condition is satisfied or the maximum number of backtracking iterations is reached.
The coefficients are then updated as
\begingroup
\setlength{\abovedisplayskip}{2pt}
\setlength{\belowdisplayskip}{2pt}
\begin{equation}
	\mathbf c_{k+1}
	=
	\mathbf c_k-\alpha_k\mathbf g_k.
	\label{eq:gd_update}
\end{equation}
\endgroup

The iteration terminates when the relative gradient norm
$\|\mathbf g_k\|_2/\max(1,\|\mathbf c_k\|_2)$ falls below
a prescribed tolerance $\varepsilon_g$, or the maximum
number of iterations $K_{\max}$ is reached. If no
acceptable step size is found within the prescribed
backtracking limit, the current estimate is retained.
Finally, the obtained frequency-domain coefficients $\hat{\mathbf c}$ are transformed back to the time domain to yield the recovered complex envelope, i.e.,
\begingroup
\setlength{\abovedisplayskip}{2pt}
\setlength{\belowdisplayskip}{2pt}
\setlength{\abovedisplayshortskip}{2pt}
\setlength{\belowdisplayshortskip}{2pt}
\begin{equation}
	\hat{\mathbf u}
	=
	\mathbf F^{\mathrm H}
	\mathbf M_{\mathbb B}\hat{\mathbf c}.
	\label{eq:gd_reconstruction}
\end{equation}
\endgroup

The initial frequency-domain coefficients are obtained via local linear approximation of the received signal. For $|u(t)|\ll A$, the nonlinear observation can be approximated as
$S(|A+u_n|)\approx S(A)+S'(A)\operatorname{Re}\{u_n\}$.
Accordingly, the linearized observation model is
$
\mathcal F(\mathbf F_{\mathbb B}\mathbf c)
\approx S(A)\mathbf{1}+
S'(A)\operatorname{Re}\{\mathbf F_{\mathbb B}\mathbf c\},
$
where $\mathbf{1}$ denotes an N-dimensional all-ones vector.
Under the assumed Gaussian noise model, the resulting linearized ML estimate has the closed-form solution
\begingroup
\setlength{\abovedisplayskip}{2pt}
\setlength{\belowdisplayskip}{2pt}
\setlength{\abovedisplayshortskip}{2pt}
\setlength{\belowdisplayshortskip}{2pt}
\begin{equation}
	\widehat{\mathbf c}_{\mathrm{lin}}
	=
	\frac{2}{S'(A)}
	\mathbf M_{\mathbb B}^{\mathrm H}\mathbf F
	\left(\mathbf y-S(A)\mathbf{1}\right).
	\label{eq:linear_initialization}
\end{equation}
\endgroup
This estimate is used to initialize the subsequent nonlinear optimization, i.e., $\mathbf c_0=\widehat{\mathbf c}_{\mathrm{lin}}$.

The detailed algorithm of PI-BLIR is summarized in Algorithm~\ref{alg:PI-BLIR}.
In terms of computational complexity, each iteration
mainly involves FFT/IFFT operations, element-wise atomic
response evaluations, and Armijo backtracking. Therefore, the overall computational complexity is $\mathcal{O}(K_{\max}J_{\max}N\log N)$, where $K_{\max}$ and $J_{\max}$ denote the maximum number of GD iterations and Armijo backtracking iterations, respectively.\footnote{The current implementation operates on a complete data frame and performs iterative reconstruction after frame acquisition, resulting in block-processing latency. A low-latency implementation will be investigated in a separate extended study.}
In particular, the proposed method requires a well-defined atomic response $S(E)$. In practice, $S(E)$ can be calibrated by continuously scanning the PD readout for known electric-field strengths. A Lorentzian function is then fitted to obtain an analytical expression for $S(E)$~\cite{ChenMinze2025}. By evaluating its slope, the optimal operating point $E_{opt}$ can be determined. 
It is worth noting that although PI-BLIR can reconstruct distorted signal waveforms, the extent of recovery is ultimately limited by atomic-response saturation.
When the input field is excessively large, atomic-response saturation occurs, causing the probe transmission to drop to zero. Consequently, the PD output contains no valid information, in which case PI-BLIR cannot recover the signal waveform.

\begin{algorithm}[!t]
	\caption{PI-BLIR Algorithm}
	\label{alg:PI-BLIR}
	\small
	\begin{algorithmic}[1]
		\REQUIRE
		Measured signal \(\mathbf y\), LO amplitude \(A\), atomic response
		\(S(\cdot)\) and its derivative \(S'(\cdot)\), frequency-domain matrices
		\(\mathbf F\) and \(\mathbf M_{\mathbb B}\),
		parameters \(\alpha_0,\mu,\rho,\varepsilon_g,K_{\max},J_{\max}\).
		
		\ENSURE Recovered complex envelope \(\hat{\mathbf u}\).
		
		\STATE Initialize \(\mathbf c_0\) via Eq.~\eqref{eq:linear_initialization}.
		
		\FOR{\(k=0,1,\ldots,K_{\max}-1\)}
		
		\STATE Compute the objective function and gradient \(\mathbf g_k\) via Eq.~\eqref{eq:gd_gradient}.
		
		\IF{$	{\left\|\mathbf g_k\right\|_2}/{\max(1,\|\mathbf c_k\|_2)}\leq\varepsilon_g$ is satisfied}
		\STATE \textbf{break}
		\ENDIF
		
		\STATE Determine step size \(\alpha_k\) via Armijo backtracking line search.
		
		\STATE Update frequency-domain coefficients: \(\mathbf c_{k+1}=\mathbf c_k-\alpha_k\mathbf g_k\).
		
		\ENDFOR
		
		\STATE Recover \(\hat{\mathbf u}\) via Eq.~(\ref{eq:gd_reconstruction}).
		
		\RETURN \(\hat{\mathbf u}\).
	\end{algorithmic}
\end{algorithm}

\section{Simulation Results}
This section evaluates PI-BLIR for communication waveform reconstruction and LDR extension using the 2C4L atomic response model. The room-temperature Cs vapor model uses $T=290~\mathrm{K}$, $L=5\times10^{-2}~\mathrm{m}$, $N_a=3.0\times10^{16}~\mathrm{m}^{-3}$, $\lambda_p=852~\mathrm{nm}$, $\lambda_c=510~\mathrm{nm}$, $P_0 = 1~\mathrm{mW}$, $\Omega_p/2\pi=0.5~\mathrm{MHz}$, and $\Omega_c/2\pi=40.284~\mathrm{MHz}$, with all central detunings set to zero. 
The remaining atomic parameters are computed using the ARC software package.
The LO amplitude is fixed at the optimal operating point.
The equivalent PD gain is set to $G_a=10~\mathrm{mA/mW}$.

For a fair comparison, three digital processing methods are evaluated under identical noise conditions:
\begin{enumerate}
	\item \textbf{$\mathbf{RAW}$}: The conventional method uses the local linear approximation at the operating point, i.e.,
	$
	u_{\mathrm{RAW}}(t)=
	\frac{y(t)-S(E_{\mathrm{LO}})}{S'(E_{\mathrm{LO}})}.
	$
	
	\item \textbf{$\mathbf{Atomic~S^{-1}}$}: Applies inverse-response $S^{-1}$ compensation to estimate the total field magnitude $r(t) = |A+u(t)|$ and subtracts the DC component of LO field to recover the signal field $u(t)$\cite{ChenMinze2025}. 
	
	\item \textbf{$\mathbf{PI\text{-}BLIR}$}: The proposed band-limited nonlinear iterative reconstruction method.
\end{enumerate}

The LDR is used to quantify the receiver's effective input range for faithful signal recovery and is defined as follows:
\begingroup
\setlength{\abovedisplayskip}{2pt}
\setlength{\belowdisplayskip}{2pt}
\setlength{\abovedisplayshortskip}{2pt}
\setlength{\belowdisplayshortskip}{2pt}
\begin{equation}
	\mathrm{LDR}
	=20\log_{10}\left(E_{\mathrm{P}_{1\mathrm{dB}}}/E_{\min}\right),
	\label{eq:sim_ldr_definition}
\end{equation}
\endgroup
where $E_{\min}$ denotes the minimum detectable input field amplitude determined by the detector noise floor, 
and $E_{\mathrm{P}_{1\mathrm{dB}}}$ is the maximum input amplitude for effective waveform recovery, determined by the 1-dB compression point ($\mathrm{P}_{1\mathrm{dB}}$) of the recovered fundamental-component amplitude. 
Thus, the effective LDR spans from the minimum detectable field to the input level at which the recovered fundamental amplitude is compressed by 1 dB. 
In the small-signal regime, all three methods exhibit negligible nonlinear distortion. Hence, under identical noise conditions, they share the same $E_{\min}$, and their LDR differences are determined solely by the maximum input amplitude $E_{\mathrm{P}_{1\mathrm{dB}}}$.

\begin{figure}[!t]
	\centering
	\includegraphics[width=0.6\columnwidth]{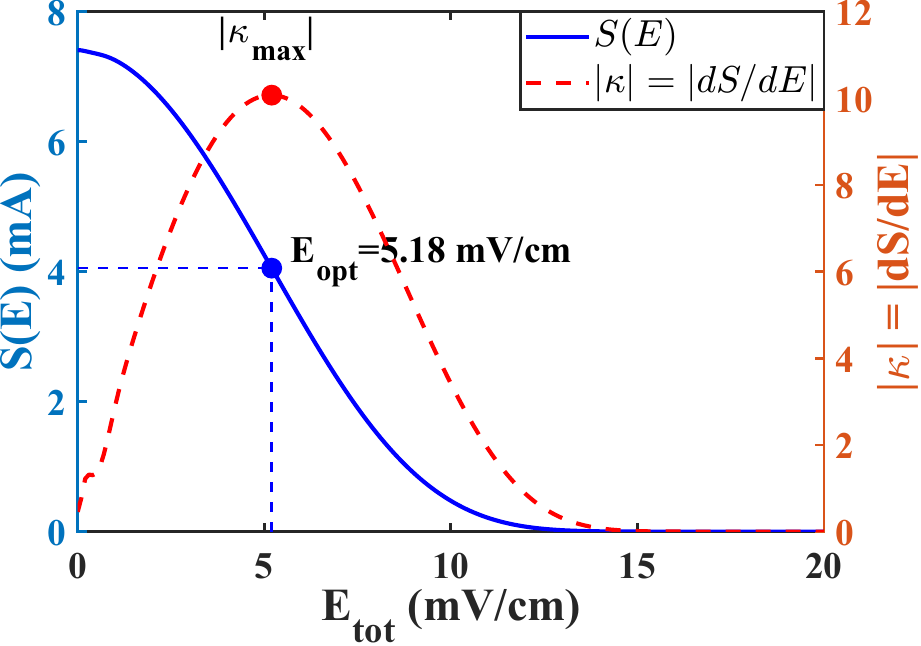}
	\caption{
		Atomic response $S(E)$ and the magnitude of its slope $|\kappa|$. 
	}
	\label{fig:sim_conditioning}
\end{figure}

Fig.~\ref{fig:sim_conditioning} shows the atomic response $S(E)$ and the magnitude
of its slope, $|\kappa|=|dS/dE|$, as functions of the total electric-field magnitude $E_{\mathrm{tot}}$.
As $E_{\mathrm{tot}}$ increases, the PD output current monotonically decreases and approaches zero due to atomic-response saturation. In contrast, $|\kappa|$ first increases and then decreases, reaching its maximum at $E_{\mathrm{opt}}=5.18~\mathrm{mV/cm}$. This point corresponds to the maximum local response sensitivity and is therefore selected as the LO operating point.

\begin{figure}[!t]
	\centering
	\includegraphics[width=0.8\columnwidth]{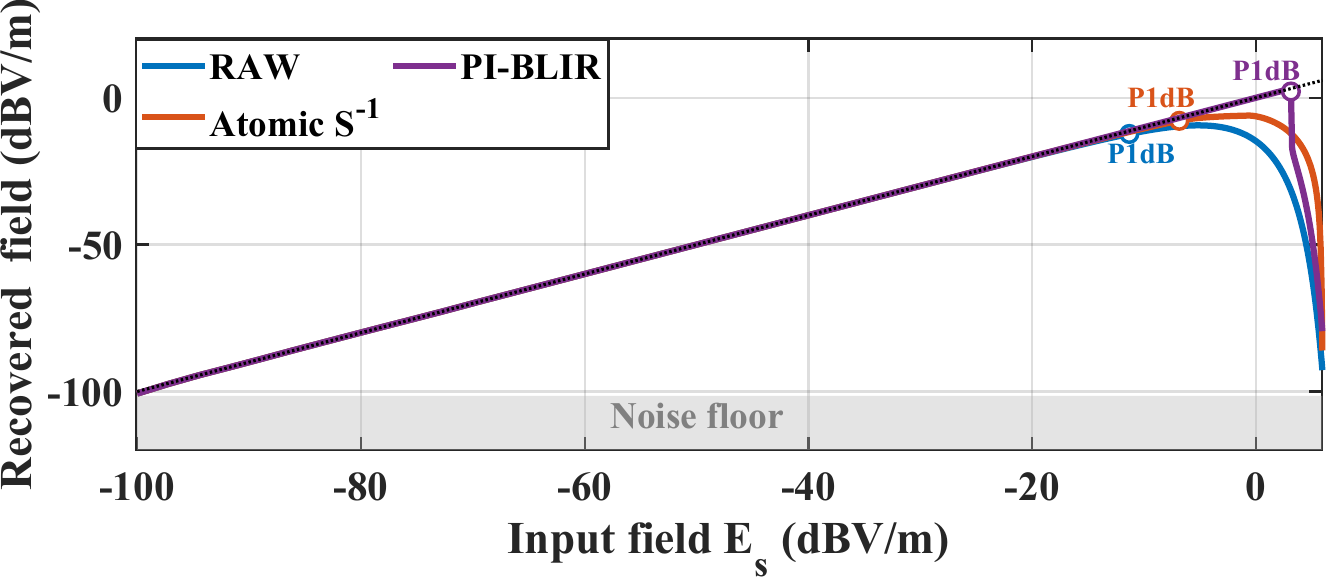}
	\caption{
		LDR Performance Comparison. 
	}
	\label{fig:LDR_comparison}
\end{figure}

To evaluate the LDR, a single-tone IF signal at $f_{\mathrm{IF}}=100~\mathrm{kHz}$ is used for testing, with the LO field fixed at $E_{\mathrm{opt}}$. The input field amplitude $E_s$ is swept to evaluate the receiver response under different input levels. 
As illustrated in Fig.~\ref{fig:LDR_comparison}, 
the vertical axis represents the electric-field amplitude corresponding to the fundamental component extracted from each reconstructed signal. 
The diagonal reference line indicates the ideal linear response with unity slope. At low input levels, all three methods closely follow this line. As the input amplitude increases, nonlinear distortion causes gain compression, with RAW reaching the 1-dB compression point first, followed by Atomic $S^{-1}$ and PI-BLIR. Specifically, PI-BLIR extends the P1dB point by 14.44 dB compared with RAW, yielding the same improvement in LDR.

\begin{figure}[!t]
	\centering
	\includegraphics[width=1\columnwidth]{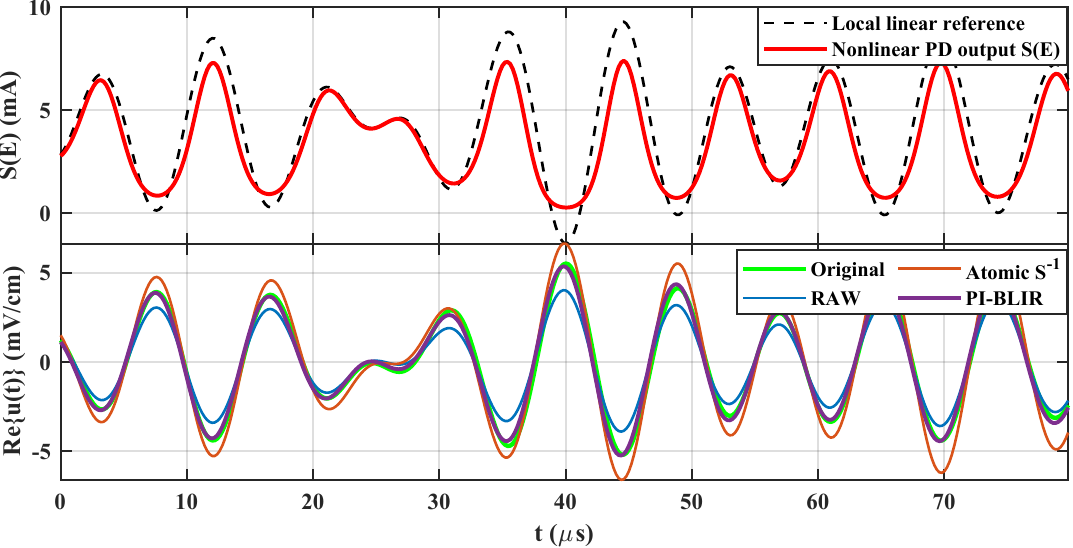}
	\caption{
		16-QAM reconstructed waveform comparison
	}
	\label{fig:waveform_comparison}
\end{figure}

\begin{figure}[!t]
	\centering
	\includegraphics[width=1\columnwidth]{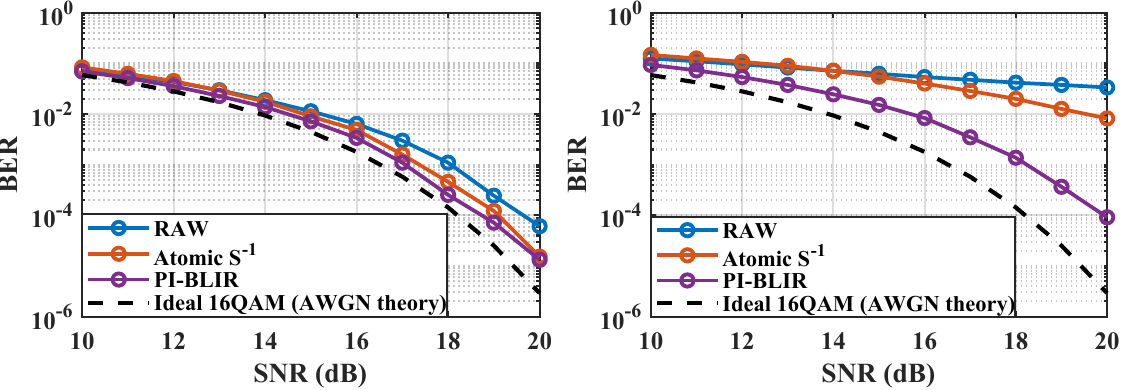}
	\caption{
		BER performance of 16-QAM signals with peak envelope amplitudes of (a) 4 mV/cm and (b) 7 mV/cm. The black dashed curves represent the ideal 16-QAM BER under AWGN.
	}
	\label{fig:BER_comparison}
\end{figure}

To evaluate the demodulation performance under different signal levels, two 16-QAM cases with peak envelope amplitudes of 4 mV/cm ($-7.96\text{ dBV/m}$) and 7 mV/cm ($-3.10\text{ dBV/m}$) are considered. 
Both cases use an IF carrier frequency of 105 kHz, a symbol rate of 40 ksym/s, and root-raised-cosine (RRC) pulse shaping with a roll-off factor of 0.2, yielding an occupied positive-frequency IF band of $\mathbb{B}=[f_L,f_H]=[81,129]$ kHz and a signal bandwidth of $48$ kHz. 
The sampling rate is 520 kHz (13 samples per symbol), and the LO field is fixed at $E_{\mathrm{opt}}$. The detector noise is modeled as additive white Gaussian noise (AWGN), with the signal-to-noise ratio (SNR) swept from 10 to 20 dB.
The SNR is defined with respect to the ideal RAW reference link, as the ratio of the average symbol energy after RRC matched filtering to the equivalent detector-noise energy after the same processing, ensuring an identical detector-noise level across all methods at each SNR.
As shown in Fig.~\ref{fig:waveform_comparison}, for a 16-QAM signal with an amplitude of 7 mV/cm at an SNR of 20 dB, the IF output waveform of $S(E)$ exhibits severe distortion compared to the local linear reference. However, reconstruction via PI-BLIR effectively mitigates this distortion and largely restores the original signal waveform.

Specifically, as shown in Fig.~\ref{fig:BER_comparison}, PI-BLIR achieves the lowest bit error rate (BER) in both cases. 
In Fig.~\ref{fig:BER_comparison}(a), where the signal amplitude is low, nonlinear distortion is weak, the three methods exhibit similar BER performance. Despite the small overall gaps, PI-BLIR yields the lowest BER, followed by Atomic $S^{-1}$ and RAW.
In Fig.~\ref{fig:BER_comparison}(b), although the BER performance of all three methods degrades in the strong-field region, PI-BLIR still maintains the best performance. 
The RAW method neglects atomic-response curvature and envelope-magnitude nonlinearity; consequently, it exhibits the worst BER performance.
For the Atomic $S^{-1}$ method, it neglects envelope-magnitude nonlinearity and becomes ill-conditioned near atomic-response saturation, leading to noise amplification. By jointly exploiting the atomic response and signal-bandwidth prior, PI-BLIR provides more effective nonlinear distortion compensation and achieves consistently better BER performance. The remaining gap between the PI-BLIR result and the ideal AWGN reference indicates residual nonlinear distortion.

\vspace{-4pt}
\section{Conclusion}
\vspace{-4pt}

In this paper, we proposed a PI-BLIR method for nonlinear distortion mitigation in Rydberg atomic receivers.
By combining the nonlinear atomic response model with prior knowledge of signal bandwidth, PI-BLIR recovers distorted signals through digital post-processing without changing the receiver's physical parameters.
Simulation results demonstrate that PI-BLIR extends the LDR by more than $14~\mathrm{dB}$ and effectively improves the BER performance of QAM demodulation in the atomic receiver.

\bibliographystyle{IEEEtran}
\bibliography{reference}
\end{document}